\documentstyle[aps,prl,epsfig]{revtex}

\begin{document}

\wideabs{
  
\title{Vortex nucleation and hysteresis phenomena in rotating Bose-Einstein condensates.}
  
\author{Juan J. Garc\'{\i}a-Ripoll and V\'{\i}ctor M. P\'erez-Garc\'{\i}a}
  
\address{Departamento de Matem\'aticas, Escuela T\'ecnica Superior de
  Ingenieros Industriales\\ Universidad de Castilla-La Mancha, 13071 Ciudad
  Real, Spain }

\date{\today}
  
\maketitle
  
\draft

%%%%%%%%%%%%%%%%%%%%%%%%%%%%%%%%%%%%%%%%%%%%%%%%%%%%%%%%%%%%%%%%%%%

\begin{abstract}
  We study the generation of vortices in rotating Bose--Einstein condensates, a
  situation which has been realized in a recent experiment (K. W.  Madison, F.
  Chevy, W.  Wohlleben, J. Dalibard, Phys. Rev.  Lett. {\bf 84} 806 (2000)). By
  combining a linear stability analysis with the global optimization of the
  nonlinear free energy functional, we study the regimes that can be reached in
  current experiments. We find a hysteresis phenomenon in the vortex nucleation
  due to the metastabilization of the vortexless condensate. We also prove that
  for a fast enough rotating trap the ground state of the condensate hosts one
  or more bent vortex lines.
\end{abstract}

\pacs{PACS number(s): 03.75.Fi, 05.30.Jp, 67.57.De, 67.57.Fg}

}

%03.75.Fi Phase coherent atomic ensemble (Bose condensation)
%05.30.Jp Boson systems
%67.57.De Superflow and hydrodynamics
%67.57.Fg Textures and vortices (superfluids)

Vortices, or vortex lines we should rather say, constitute the most relevant
topological defect in Physics. They consist on a twist of the phase of a wave
function around an open line and they are typically associated to a rotation of
a fluid whatever the fluid is made of (real fluids, optical fluids, quantum
fluids, ...)  \cite{vortices_real}.  Vortices are one of the means by which
quantum systems acquire angular momentum and react to perturbations of the
environment. They have already been predicted, observed and studied in the
superfluid phase of $^4$He and are indeed known to be the key to some important
processes in these systems, such as dissipation, moments of inertia and
breakdown of superfluidity. This is why extensive research on vortex
generation, stability and dynamics has been conducted in the field of
Bose-Einstein condensation (BEC) in the last years
\cite{vortices-BEC,Rokshar,us-vortex,us-JILA}.

Vortices and other defects usually involve more energy than other equilibrium
states, such as convex nodeless states. Therefore, in order to produce a vortex
in a condensate, one must induce a transition from a uncondensed or condensed
convex cloud to the desired state, by means of an external action such as a
change of the confining potential.

For instance, in recent experiments performed by a group from the \'Ecole
Normale Sup\'erieure (ENS) \cite{ENS}, vortices are created by rotating an
elongated trap which is slightlty deformed along its transverse dimensions.
After this preparation, the trap is switched off, and the condensate expands
until vortices become directly observable.

In these experiments there are several controversial points which are: (i)
Vortices first nucleate at a trap rotation speed or \emph{critical frequency},
$\Omega_1$, larger than that required to stabilize a vortex line in the
Thomas-Fermi theory, (ii) when seen from the top vortex cores seem partially
filled and (iii) after the nucleation of the first vortex the angular momentum
grows continuously with the rotation speed (not only with discontinuous jumps).

In this paper we reach a global view of the transitions between equilibrium
states which are induced by the trap rotation, plus a simple explanation of
the most controversial points. Our study consists of two parts.  First we focus
on the analysis of stationary states and discuss the stability properties of
the simplest solutions, i.e. straight vortices, which allows us to get some
insight on the problem. Our main result is that the metastabilization of a
vortexless state induces hysteresis in the vortex nucleation
process. In other words, the trap rotation speed must exceed a critical value,
$\Omega_M$, to produce vortices, but that speed is well beyond the values,
$\Omega_1, \Omega_2, \ldots$, which are required to make vortices energetically
favorable.  In the second part we find numerically the actual ground states of
the condensate for different angular speeds. We will show that once the first
vortex is nucleated the angular momentum grows almost continuously with the
speed of the trap, while the ground state (GS) mutates into different deformed
states.

Several other works have contributed to the theoretical description of the
experimental ``anomalies''. First, Ref. in \cite{Fetter-new} the dynamics of vortex
lines is studied analytically, and different modes are obtained which reflect
the transverse tension of vortex lines.  Next, in Ref. \cite{Feder} the existence of
these modes is confirmed numerically a relation between these modes and the
large value of the critical angular speed of the ENS experiments is proposed.
Finally in Ref. \cite{Stringari} the authors derive a condition for
the efficiency of external perturbations when trying to induce a
mechanical response in the condensate.

\emph{The model.-} For most of current experiments it is an accurate
approximation to use a zero temperature many-body theory of the condensate. In
that limit the whole condensate is described by a single wavefunction
$\psi({\bf r},t)$ ruled by a Gross-Pitaevskii equation (GPE) \cite{Dalfovo}.

In Ref. \cite{ENS} the trap is initially harmonic with axial symmetry, but then a
laser is applied which deforms it and makes it rotate with uniform angular
speed $\Omega$. On the mobile reference frame which rotates with the trap, the
experiment is modeled by a modified GPE
\begin{equation}
\label{GPE-rot}
i\frac{\partial \psi}{\partial t}  =\left[ -\frac{1}{2}\triangle +
V_0({\bf r}) + \frac{g}{2}\left| \psi \right| ^{2}
-\Omega L_{z}\right] \psi.
\end{equation}
Here $L_{z}=i\left(x\partial_y-y\partial_x\right)$ is the hermitian
operator that represents the angular momentum along the z-axis and
the effective trapping potential in given by
\begin{equation}
V_0({\bf r}) =
\frac{1}{2}\omega_\perp^2(1-\varepsilon)x^2+
\frac{1}{2}\omega_\perp^2(1+\varepsilon)y^2+
\frac{1}{2}\omega_\perp^2\gamma^2z^2.
\end{equation}

In Eq. (\ref{GPE-rot}) we have applied a convenient adimensionalization which
uses the harmonic oscillator length, $a_\perp=\sqrt{\hbar/m_{Rb}\omega_\perp}$
and period, $\tau =\omega_\perp^{-1}.$ With these units the nonlinear parameter
becomes $g=4\pi a_{\text{S}}/a_\perp$, where $a_{\text{S}}\simeq 5.5\,
\textrm{nm}$ is the scattering length for $^{87}$Rb, the gas used in Ref.
\cite{ENS}. Following the experiment we will take $\omega_z=2\pi \times 11.6\,
\textrm{Hz}$ and $\omega_\perp=2\pi \times 232\, \textrm{Hz}$, and we will use
$Ng = 10000$, which corresponds to a few times $10^5$ Rb atoms.  For the small
transverse deformation of the trap we have tried $\varepsilon = 0.0,0.03\text{
  and }0.06$ ($\varepsilon=0.03$ is the closest one to the actual experiment
\cite{ENS}).

The norm, $N[\psi]=\int \left| \psi \right| ^{2}d{\bf r}$, which is related to
the number of bosons in the condensate, and the energy
\begin{eqnarray}
E[\psi ] & = & \int \bar{\psi }\left[ -\frac{1}{2}\triangle +V_{0}\left( \mathbf{r}\right) 
+\frac{g}{2}\left| \psi \right| ^{2}-\Omega L_{z}\right] \psi d{\bf r} \nonumber \\
& = & E_0(\psi) - \Omega  L_z(\psi). \label{energy}
\end{eqnarray}
are conserved quantities.  Each stationary solution of the GPE of the form
$\psi _{\mu}({\bf r},t) =e^{-i\mu t}\phi({\bf r}),$ is a critical point of the
energy (\ref{energy}) with the  norm constraint, $\left. \frac{\partial
    E}{\partial \psi }\right| _N[\psi_\mu]=0$.

\emph{Critical frequencies.-} Each state of the system is a ``point'' in the
infinite--dimensional functional space on which the energy $E_0(\psi)$ and the
angular momentum $L_z(\psi)$ are defined. In principle, due to the conservation
of the energy, and due to the fact that dissipation at most drains energy out
of the system, the minima of $E_0(\psi)$ are associated to stable states.
Around these minima, the ``motion'' of the system is confined by energy
barriers.  However, rotation involves a pointwise change of the height of these
infinite dimensional ``surfaces'' of energy, which move from $E_0(\psi)$ to
$E_0(\psi)-\Omega L_z(\psi)$. This shift, which is stationary for each
dynamical configuration of the trap adopted, can turn stable local minima into
saddle points and open paths for the evolution of the condensate from simple nodeless state
to states with one or more vortices.

Let us consider the case of an axially symmetric trap. These traps admit
axially symmetric solutions with either no vortices, $\psi_0$, or with one
centered vortex of integer charge, $\psi_m \propto \exp\{\hbox{i}m
\arctan(y/x)\}$. In nonrotating traps, the vortexless state is actually the one
with less energy. As the rotation speed is increased, the energies of the states
with positive vorticity are all shifted down, $E_m(\Omega) = E_m(0) - m\Omega$,
so that when $\Omega > \Omega_m = \frac{E_m(0)-E_0(\Omega)}{m}, m=1,2,...,$ the
state $\psi_m$, i.~e. a symmetric and straight vortex with vorticity $m$, is
energetically more favorable than $\psi_0$. The values $\Omega_m$ are thus
first estimates for the critical frequencies at which states with vorticity $m$
could be created.

We must note that the condition $\Omega > \Omega_m$, does not imply that
$\psi_m$ is a global minimum but only that $E(\psi_0) > E(\psi_m)$. It is feasible
for other states with the same vorticity to have less energy than the straight
vortices. That is, another state $\tilde{\psi}_m$ might exist such that
$E(\tilde{\psi}_m) < E(\psi_m)$. This is known to be the case for $\psi_2$
which is energetically less favorable than a pair of $m=1$ vortices
\cite{Rokshar,us-vortex}.

In fact, for an axisymmetric vortex $\psi_m$ to be stable, it should be at
least a local minimum of $E(\psi)$. One way of studying the local stability is
by using linear stability analysis, i.~e. we linearize the energy functional
around any state $\delta({\bf r})$
\begin{equation}
E[\psi+\delta] = E[\psi] + E'[\psi](\delta) + E''[\psi](\delta,\delta)+
{\cal O}(\vert\delta\vert^3)
\end{equation}
and find a quadratic expansion
\begin{eqnarray}
E''[\psi](\delta,\delta) &=& \sum_{m,j} \lambda_{m,j}(\Omega;\psi) |\delta_{m,j}|^2,
\end{eqnarray}
in which the perturbation is expanded on a suitable basis
$\phi_{m,j}(\rho,z)e^{im\theta}$ \cite{us-vortex} as $\delta({\bf r}) =
\sum_{m,j}\delta_{m,j}\phi_{m,j}(\rho,z)e^{im\theta}$.  Following this
procedure we are able to study the curvatures of the surface $(\psi,E[\psi])$
around that state ---namely, the eigenvalues $\lambda_{m,j}(\Omega;\psi) =
\lambda_{m,j}(0;\psi)-m\Omega$---. This analysis allows us to decide when 
the operator $E''[\psi_m]$ is positive definite and consequently when a particular
$\psi_m$ becomes a local minimum.  This happens for a critical frequency that
will be denoted by
\begin{equation}
  \bar{\Omega}_{\ell} =-\min_{j,m} \frac{\lambda_{m,j}(0;\psi_{\ell})}{m-{\ell}}.
\end{equation}
For instance, when $\Omega = \bar{\Omega}_1$ the straight vortex becomes a
local minimum.

To generate a condensate with vorticity two conditions must be satisfied. It is
clear that the trap must rotate fast enough to make a vortex state
energetically favorable, but it is also necessary that an energy decreasing
path exists from $\psi_0$ to the vortex state, otherwise the transition could
be inhibited due to the energy barrier around $\psi_0$ making this state
metastable. Mathematically, there should exist a perturbation (a ``direction''
in the phase space), $\phi_{m,i}$, such that $\Omega >
\lambda_{m,i}(0;\psi_0)/m$, a condition which looks similar to the bounds of
Ref. \cite{Stringari}.  We will define a destabilization frequency, given by
\begin{equation}
  \Omega_M = \min_{m,j} \left\{\frac{\lambda_{m,j}(0;\psi_0)}{m}\right\}.
\end{equation}
Within the interval $\Omega \in [0,\Omega_M)$ a condensate in a vortexless
state cannot be forced to acquire angular momentum unless some additional
energy is pumped in and the nodeless  state $\psi_0$ could be metastable.

{\em Results for symmetric vortices.-} One could be tempted to think that the
real situation is the simplest one, i.~e. that by increasing $\Omega$ one
should reach a point $\Omega_1$ in which $E(\psi_0) > E(\psi_1)$ and at the
same time $\Omega_1 \geq \bar{\Omega}_1, \Omega_M$. In this situation $\psi_1$
would become also a local minimum $(\bar{\Omega}_1)$ and the core state would
loose its stability $(\Omega_M)$. However this is not the case, and the
situation is more complicated as we will see below.

\begin{figure}
\begin{center}
  \epsfig{file=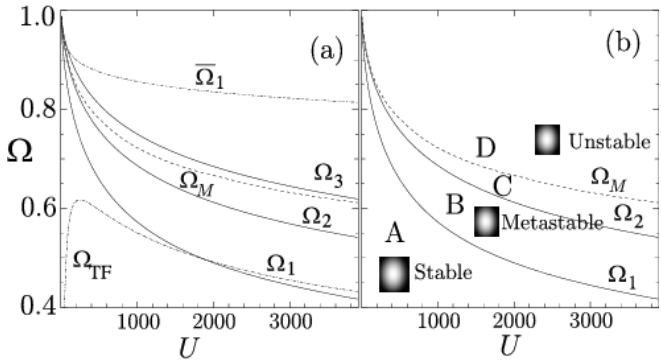,width=\linewidth}
\end{center}
\caption{\label{fig-omega}
  (a) Critical angular speeds for an elongated condensate
  ($\omega_z/\omega_\perp = 18.7$) at different nonlinearities ($U = gN 
  = 4\pi N a_{Rb}/a_\perp$). We plot the Thomas-Fermi estimate for the energy
  difference between a core and vortex state ($\Omega_{TF}$, dash--dot line)
  the speeds $\Omega_1,\Omega_2$ and $\Omega_3$ at which vortices with
  charge $m=1,2,3$ become energetically favorable (solid lines), and
  the angular speed at which $\psi_0$ (vortexless state) becomes a saddle point
  ($\Omega_M$, dashed line). (b) Regions with different phenomenology: from A
  to C there is an energy barrier surrounding the vortexless state; in A, B and
  C the ground state has vorticity equal to 0, 1 and 2 respectively; in D the
  energy barrier around the core state vanishes and more and more vortices
  become feasible.}
\end{figure}

With the previous definitions in mind we have computed numerically the states
$\psi_0, \psi_1, ...$ as well as the different frequencies $\Omega_1, \Omega_2,
\Omega_3,\bar{\Omega}_1$ and $\Omega_M$.  The results are plotted in Fig.
\ref{fig-omega} for a very elongated trap such as those used in the ENS
experiment \cite{ENS}.

There are two relevant conclusions that may be obtained from this picture. The
first one is that $\bar{\Omega}_1 > \Omega_1,\Omega_2$ which means that the
ground state of the system may never be a symmetric vortex line. In other
words, should it be energetically favorable for the ground state to acquire some
vorticity, it will never be by means of straight vortex lines, but with some
other structure.  From a practical point of view, the large difference between
$\Omega_1$ and $\bar{\Omega}_1$ also implies that the ENS experiment is working
on a regime in which straight vortex lines are very hard to obtain. What are
the new vortex structures which appear instead above $\Omega_1$ is a point we will discuss
later.

The second and most relevant feature is that the vortexless state, $\psi_0$,
remains a local minimum up to a speed, $\Omega_M$, which is much larger than
the value at which vortices become energetically favorable (Regions B and C in
Fig.  \ref{fig-omega}(b)). Before reaching $\Omega = \Omega_M$ it is
energetically expensive to introduce angular momentum into the condensate, since an
energy barrier must be surpassed.  For $\Omega > \Omega_M$ states with one or
more vortices become feasible at the same time. This persistent metastability
of $\psi_0$ should be responsible for the experimental observation that a
single angular speed is capable of producing states with a different number of
vortex lines \cite{ENS}.

\begin{figure}
\begin{center}
  \epsfig{file=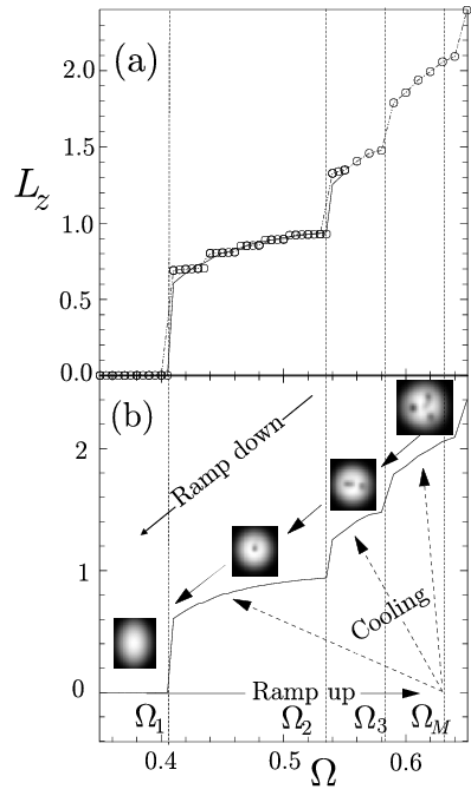,width=0.7\linewidth}
\end{center}
\caption{\label{fig-ens}
  (a) Angular momentum, $L_z$, of the ground state of the energy functional
  $E(\psi)$ [Eq. (\ref{energy})] as a function of $\Omega$. Circles correspond
  to solutions over a grid with $32\times 32\times 64$ Fourier modes for
  $\varepsilon=0,0.03,0.06$. The solid line corresponds to the solutions on a grid with
  $64\times 64\times 128$ modes and $\varepsilon = 0.03$. (b) Schematic
  picture of the angular momentum of the ground state versus angular speed,
  with a graphical description of the hysteresis mechanism. All figures are adimensional.}
\end{figure}

Our findings imply that once a vortex state is reached, the rotation speed may
be ramped down and vortices should remain stable for some range of $\Omega$
values. For instance, if after reaching $\Omega_M \simeq 0.63\omega_{\perp}$
for $U = gN = 3000$ one gets a state with a vortex, then the rotation speed may
be ramped down until $\Omega_1 \simeq 0.46\omega_{\perp}$ giving rise to a
hysteresis phenomenon as it is graphically illustrated in Fig.
\ref{fig-ens}(b).

\emph{True ground states.-} For $\Omega > \Omega_M$, or if the the energy
barrier which surrounds the core state is overstepped, a condensate produced by
means of evaporative cooling should correspond to the absolute minimum of the
energy in the configuration space. We have worked numerically with the energy
functional [Eq. (\ref{energy})] in three spatial dimensions using a technique
known as Sobolev gradients to find the ground states subject to some reasonable
constraints ---i.~e. the norm and angular speed--- (simple minimization methods
do not work for this problem).  The details of the procedure are given in Ref.
\cite{Sobolev}. We have applied this method on a Fourier basis with $32\times
32\times 64$ modes (enough for plotting purposes) and with $64\times 64\times
128$ modes (which is needed to lower the error in $L_z$ below $1\%$.)

\begin{figure}
\begin{center}
  \epsfig{file=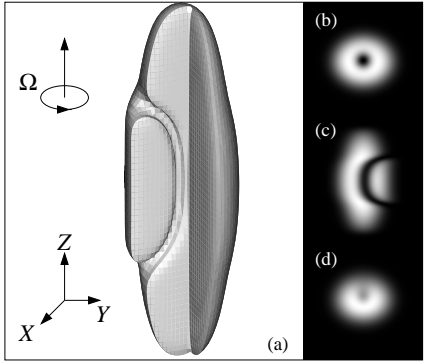,width=0.65\linewidth}
\end{center}
\caption{\label{agujero}
  Shape of the ground state of the Hamiltonian (\ref{energy}) for $\Omega =
  0.75 \omega_{\perp}, \gamma = 18.7$, $Ng = 1000$ and $\varepsilon=0.06$.  (a)
  Transverse section. (b) Density plot. One quarter of the condensate has been
  removed to allow the direct observation of the inner structure of the vortex
  line.}
\end{figure}

 Figs. \ref{fig-ens} and
\ref{agujero} summarize our results. In Fig. \ref{agujero}(a) we see that the ground state
acquires vorticity when $\Omega$ is smaller than the core--state
destabilization speed and the experimental values found in \cite{ENS},
in agreement with the predictions of the preceding paragraphs. Our
numerical method also solves the question posed before: if the ground state
must have some vorticity and it is not a straight vortex, what is its shape?
In Fig. \ref{agujero} we see that vortices are nucleated with deformed shapes.
The combination of the sudden destabilization of the vortexless state, and the
longitudinal deformations of vortex lines is enough to explain why vortex
lines seem partially filled and why the angular momentum evolves almost
continuously above $\Omega_M$. More details on the structure of bent vortices
will be given elsewhere \cite{elsewhere}.

\emph{Discussion and conclusions.-} We have used the GPE to study the
nucleation of vortices in an elongated trap similar to that used at the ENS
\cite{ENS}. Our main prediction is that in such elongated traps there exists a
mechanism which prevents the nucleation of vortices unless the condensate
rotates much faster than the speed required to make one or more vortices
energetically stable (metastabilization of $\psi_0$).

We predict that in such elongated traps, straight vortex lines are locally
stable only for very large angular speeds ---well beyond the values at which
higher vorticites become preferable---.  And we find the longitudinal
deformation of these topological defects to be responsible for an almost
continuous growth of the angular momentum with respect to the angular speed.

It is difficult to observe the bending of vortex lines in current experimental
setups since condensates are expanded too much along their transverse
dimensions.  Nevertheless, other predictions of this paper such as the
metastabilization phenomenon should be easily testable.  First, before the
destabilization of the vortexless state, $\Omega_M$, the vortex nucleation
process is prevented, and the condensate may only adapt to the rotation speed
by means of transverse deformations which are described in \cite{us-asym}. And
second, a more controlled nucleation of vortices is possible by ramping up the
trap beyond $\Omega_M$ and then carefully slowing down the condensate to some
point above $\Omega_1$. An expansion of such cloud should lead to the
observation of vortices for much lower values of $\Omega$ than those reported
in Ref.  \cite{ENS}.

This work has been partially supported by CICYT under grant PB96-0534.

\end{document}